\documentclass[11pt]{article}

\usepackage{amssymb}
\usepackage{amsmath}
\usepackage{cite}
\usepackage{epsf}

\newcommand{\sfrac}[2]{{\textstyle\frac{#1}{#2}}}
\newcommand{\tomega}{{\tilde \omega}}

\begin{document}
%%%%%%%%%%%%%%%%%%%%%%%%%%%%%%%%%%%%%%%%%%%%%%%%%%%%%%%%%%%%%%%%%%%%%%%%%%%%%%%%

\begin{titlepage}
%%%%%%%%%%%%%%%%%%%%%%%%%%%%%%%%%%%%%%%%%%%%%%%%%%%%%%%%%%%%%%%%%%%%%%%%%%%%%%%
\begin{flushright}
LBNL-45490\\
UCB-PTH-00/11\\
{\tt hep-th/0004133} \\
\end{flushright}

%%%%%%%% The Bible of Self-Tuners %%%%%%%%%%%
\vskip.5cm
\begin{center}
{\huge{\bf General Properties of the Self-tuning}}
\vskip.2cm
{\huge{\bf  Domain Wall Approach to the}}
\vskip.2cm
{\huge{\bf  Cosmological Constant Problem}}
\end{center}
\vskip0.2cm

\begin{center}
{\sc Csaba Cs\'aki} $^{a,}$\footnote{J. Robert Oppenheimer 
fellow.}, {\sc Joshua Erlich} $^{a}$,
{\sc Christophe Grojean} $^{b,c}$ and \\
{\sc Timothy J. Hollowood} $^{a,d}$
\end{center}
\vskip 10pt

\begin{center}
$^{a}$ {\it Theory Division T--8,
Los Alamos National Laboratory, Los Alamos, NM 87545, USA} \\ 
\vspace*{0.1cm}
$^{b}$ {\it Department of Physics,
University of California, Berkeley, CA 94720, USA} \\ \vspace*{0.1cm}
$^{c}$ {\it Theoretical Physics Group,
Lawrence Berkeley National Laboratory, \\ Berkeley, CA 94720, USA} \\ 
\vspace*{0.1cm}
$^d$ {\it Department of Physics,
University of Wales Swansea, SA2 8PP, UK}
\end{center}

\vglue 0.3truecm

\begin{abstract}
\vskip 3pt

\noindent
We study the dynamics of brane worlds coupled to a scalar field and gravity, 
and find that self-tuning of the cosmological constant is generic 
in theories with at most two branes or
a single brane with orbifold boundary conditions.  
We demonstrate that singularities are generic in the self-tuned
solutions compatible with localized gravity on the brane: we
show that localized gravity
with an infinitely large extra dimension is only consistent with particular
fine-tuned values
of the brane tension. The number of allowed brane tension values is 
related
to the number of negative stationary points of the scalar bulk potential and, 
in the
case of an oscillatory potential, the brane tension 
for which gravity is localized without singularities
is quantized.
We also examine a resolution of the
singularities, and find that 
fine-tuning is generically re-introduced at the singularities in order to
retain a static solution.  However,
we speculate that the presence of additional fields may restore self-tuning.

\end{abstract}

\end{titlepage}

%%%%%%%%%%%%%%%%%%%%%%%%%%%%%%%%%%%%%%%%%%%%%%%%%%%%%%%%%%%%%%%%%%%%%%%%%%%%%%%
\section{Introduction}
\setcounter{equation}{0}
\setcounter{footnote}{0}

There has been renewed interest in Kaluza--Klein theories over the past two
years, mainly due to the realization that localization of matter \cite{large}
and  localization of gravity \cite{RS} may drastically change the commonly
assumed properties of such models. These theories clearly open new
approaches to the cosmological constant problem\footnote{The cosmological
constant problem concerns the brane cosmological constant that governs
the expansion of our universe and it has to be
distinguished to the vacuum energy density, or tension, on the brane:
in particular it is possible, for a non-vanishing tension, to find solutions
to the Einstein equations with a flat Minkowski metric on the brane
in which case the brane cosmological constant is zero.}, since using extra
dimensions the no-go theorem of Weinberg for adjustment mechanisms 
\cite{Weinberg} may be circumvented. A particularly simple scenario has been
recently suggested in \cite{ADKS,KSS} to at least improve on the 
cosmological constant problem (see also
\cite{brane-cc-old} for an earlier mechanism involving
extra dimensions), by proposing a mechanism for the cancellation
of all order Standard Model (SM) loop contributions and the leading (tree-level)
gravity contribution to the effective 4D cosmological constant,
\begin{equation}
\underbrace{\mathcal{O} (M^4)}_{\mbox{\small tree-level}} +
\underbrace{\mathcal{O} (T_{br})}_{\mbox{\small SM loops}} +
\underbrace{\mathcal{O} (T_{br}^2 M^{-4}) + \ldots}_{\mbox{\small higher order}}
\end{equation}
where $M$ is the fundamental scale of gravity, the Planck scale in the bulk
for instance.
The mechanism is based on a single 3-brane (to which the SM fields are
localized) embedded into a 4+1 dimensional space-time. The essential new
ingredient is a bulk scalar field $\phi$, which is coupled to the brane
tension $T_{br}$ (assumed to be the full loop-corrected SM vacuum energy density).
The authors of \cite{ADKS,KSS} then showed that one can find static solutions
to the classical equations of motion for a vanishing bulk potential of the 
scalar field, but for arbitrary values of the brane tension $T_{br}$.
However, all the solutions found in \cite{ADKS,KSS} which localize gravity
involve a naked
space-time singularity, which has been interpreted as the boundary of the
extra dimension (see also \cite{horizon} for a discussion on singularity
in a brane-world context).
Since the bulk is effectively compactified, we have
to worry whether the size of the extra dimension is compatible with our
phenomenological knowledge of gravity which has been tested up to millimeter
distances. The proper distance from the brane to the singularity
is found to be $y_c\sim \kappa_5^{-2}\, T_{br}^{-1}$ where
$M_5= \kappa_5^{-2/3}$ is the 5D Planck scale, which is 
related to the 4D Planck scale,
$M_4=\kappa_4^{-1}$, by $T_{br}=\kappa_4^2/\kappa_5^4$.
If the tension, associated to the vacuum energy of the SM fields, is of the order
of the electroweak scale, $T_{br} \sim \mbox{TeV}^4$,
we naturally obtain\footnote{The fact that the electroweak scale on the brane is
five orders of magnitude below the fundamental scale, $M_5$, is the gauge hierarchy
problem in this model and a mechanism, such as low energy supersymmetry for instance,
is needed to cancel the quadratic divergences which should bring $T_{br}$ near
$M_5^4$.}
\begin{equation}
M_5 \sim 10^8 \, \mbox{GeV } \ \ \
\mbox{and } \ \ \ y_c \sim 1 \, \mbox{mm}
\end{equation}
which is phenomenologically safe. Finally, while the SM contribution
to the 4D cosmological constant would be of the order of $10^{-64}\, M_4^4$,
the self-tuning mechanism cancels this term.
However, the next possible term is
of the order of $10^{-84}\,M_4^4$, which is still forty orders
of magnitude too large. It is worth noticing that the self-tuning mechanism
eliminates twenty orders of magnitude and thus should be considered on 
equal 
footing to the Randall--Sundrum solution to the hierarchy problem.

The necessity of singularities in the bulk is here only dictated by
the phenomenological requirement of localized gravity. However
it complements the singularity theorem of Hawking and Penrose under which 
generic
initial conditions lead to singular solutions of Einstein's equations 
\cite{HP}.

Several works \cite{brane-cc-new,deAlwis} have studied the self-tuning mechanism.
In this paper, we examine the general properties of
the solutions to the coupled 5D scalar-gravity
system in detail. First, we investigate the solutions in the presence
of a general bulk potential for the scalar field.
The motivation to include a scalar bulk potential
are twofold: {\it (i)} to give a mass to the scalar field and thus evade
the cosmological problems associated to a massless scalar field coupled to 
gravity
and/or  those associated to an unstabilized extra-dimension;
{\it (ii)} to overcome the singularity problem by considering a more general
bulk geometry.
We explain that the self-tuning behavior is expected to
occur for a generic bulk potential. However, in the generic solutions
(except for a vanishing bulk potential) self-tuning does have the more
restricted interpretation that there is at least a finite region for the
brane tension for which static solutions can be found (but not necessarily
for all values of $T_{br}$). More precisely, Standard Model corrections are
expected to occur at the weak scale, whereas self-tuning works up to the
larger 5D Planck scale.  

After presenting our general results and methods 
on the 
solutions in the presence of a generic bulk potential we ask the 
following question:
are there bulk potentials such that the self-tuning solutions 
avoid naked space-time
singularities for a range of values of the brane tension in 
such a way, that gravity is localized to the brane (so as to reproduce 4D 
gravity for the observer on the brane). We show that for the 5D scalar-gravity
system the only possibility for such solutions is 
when space-time asymptotes to five dimensional anti-de Sitter space 
(AdS$_5$) far from the brane (therefore producing solutions
of the sort considered in \cite{DZF}). After a careful analysis we show that
such solutions are always fine-tuned; that is, they occur only for particular
isolated values of the brane tension $T_{br}$. We show that the number
of allowed brane tensions is related to the number of negative stationary
points of the bulk potential, which appears as a maximal index.
In the case of an oscillatory bulk potential, we thus obtain a quantization
of the brane tension as in a brane world construction from supergravity
\cite{CGS3}.

The paper is organized as follows: in Section 2 we present our method
for searching for solutions of the coupled scalar-gravity system using
the superpotential formalism of \cite{DZF,W,Gremm,CEHS}. The advantage
of this method is that it results in first order ordinary differential 
equations, and all the degrees of freedom are transparent without having to
make a particular ansatz for the solution. In Section 3 we present a class
of exactly solvable systems, given by an exponential bulk potential,
which includes all the models presented in \cite{ADKS,KSS}. We analyze
these models in detail and find that all of these solutions necessarily
involve a naked singularity or lead to an infinite Planck scale on the brane.
We also find that, by relaxing an ansatz for solutions made in \cite{KSS},
the exponential bulk potential is in fact self-tuning.
In Section 4 we compare a perturbative
and a numerical method to solve the equations for the most general 
bulk potentials, and present an example for both methods using
the exactly solvable cases of Section 3. In Section 5 we present a
no-go theorem for self-tuning branes that would localize gravity without
singularities. In Section 
6 we comment on the effect of resolution of the singularities on self-tuning. 
We conclude in Section 7.

%%%%%%%%%%%%%%%%%%%%%%%%%%%%%%%%%%%%%%%%%%%%%%%%%%%%%%%%%%%%%%%%%%%%%%%%%%%%%%%
\section{Self-tuning and scalar fields}
\label{sec:self-tune}
\setcounter{equation}{0}
\setcounter{footnote}{0}

In this section we demonstrate that the 4D cosmological constant of branes
coupled to scalar fields is generically self-tuned to zero,
as in \cite{ADKS,KSS}.
We begin with the action for a brane coupled to gravity and a real scalar 
field (such dilatonic domain walls have been extensively studied
from a supersymmetric point of view \cite{cvetic}, a new ingredient here
is the localization of SM fields)\footnote{Our conventions
correspond to a mostly positive Lorentzian
signature
$(-+\ldots +)$ and the definition of the curvature in terms of the metric is
such that a Euclidean sphere has  positive curvature. Bulk indices will be denoted
by capital Latin indices and brane indices by Greek indices.},
\begin{equation}
	\label{action}
S=\int d^Dx \,\sfrac{1}{2\kappa^2_D} \sqrt{|G|}
\left[ R-\sfrac{D-2}{D-1}
\left(\partial_M\phi\partial^M\phi+V[\phi]\right)\right]-
\int d^{D-1}x\,\sqrt{|g|}\,\sfrac{D-2}{(D-1)\kappa^2_D}\,f[\phi]\,T,
\end{equation}
where $G_{MN}$ is the $D$-dimensional metric and $g_{\mu\nu}$ is the
pullback of the metric onto the flat domain wall at $x^{D-1}\equiv y=0$.
For now we will not be concerned with the origin of the coupling 
$f[\phi]$ to the
brane, and allow it to be arbitrary; $T$ is the non-canonically normalized
brane tension and includes standard model vacuum contributions.
We look for static solutions with the metric ansatz\footnote{We will not
look for non-flat solution on the brane such as de Sitter or anti--de Sitter
4D configurations. When they exist simultaneously with flat Minkowskian solutions,
it is a dynamical question to know which solution is preferred by stability. 
A nice feature of the case with vanishing bulk potential is that the
flat solutions are the only 4D maximally symmetric solutions \cite{ADKS}.},
\begin{equation}
	\label{metric}
ds^2=e^{-2A(y)/(D-1)} dx^2_{D-1} + dy^2.
\end{equation}
The unconventional normalization of the action \eqref{action} and the warp
factor in \eqref{metric} will be convenient in what follows.

The equations of motion which follow from  the action \eqref{action}
with the ansatz \eqref{metric} are \cite{DZF,CEHS},
\begin{eqnarray}
	\label{ein1}
&&
A''(y)  =  \phi'(y)^2 + f[\phi(y)]\, T\, \delta(y),
\\
	\label{ein2}
&&
A'(y)^2 = \phi'(y)^2-V[\phi(y)],
\\
	\label{ein3}
&&
\phi''(y)-A'(y)\phi'(y) =
\sfrac{1}{2}\,\frac{\partial V[\phi]}{\partial\phi}+
\frac{\partial f[\phi]}{\partial\phi}\, T\, \delta(y),
\end{eqnarray}
where the primes denote derivatives with respect to $y$.
If $\phi(y)$ is monotonic in the bulk then the bulk equations of motion can be
written in a first order form \cite{DZF,W,CEHS} introducing an auxiliary
field $W[\phi]$,
\begin{eqnarray}
	\label{V}
V[\phi]&=&
\left(\frac{\partial W[\phi]}{\partial\phi}\right)^2-W[\phi]^2 ,
\\
	\label{phi'}
\phi'(y)&=&\frac{\partial W[\phi(y)]}{\partial\phi(y)},
\\
	\label{A'}
A'(y)&=&W[\phi(y)] .
\end{eqnarray}
Because of the relation \eqref{V}, $W[\phi]$ will be called {\it superpotential}
even though no supersymmetry is involved.
We will use the first order formalism in order to construct solutions in the bulk
on both sides of the brane: $W_\pm$, $\phi_\pm$ and $A_\pm$
 on the right ($+$) and left ($-$) hand side. But
the equations of motion \eqref{V}-\eqref{A'} must then be supplemented by 
boundary conditions at the brane.
The boundary conditions due to the delta-function terms in \eqref{ein1} and
\eqref{ein3} can be written,
\begin{eqnarray}
&&
\Delta \phi' =
\Delta \frac{\partial W[\phi(y)]}{\partial\phi(y)} =
T\, \frac{\partial f[\phi(y)]}{\partial\phi(y)}{}\Big|_{y=0}
\nonumber \\
&&
\Delta A' =
\Delta W[\phi(y)]= T\, {f[\phi(y)]}\Big|_{y=0},
\label{bc} \end{eqnarray}
where $\Delta F$ indicates the jump of a discontinuous function $F$ at
$y=0$, $F(0^+)-F(0^-)$.
In addition, $\phi(y)$ and $A(y)$ must be continuous across the boundary.
Hence, there are four boundary conditions at the brane.  

A count of free
parameters in the solutions to the equations of motion immediately demonstrates
that given $f[\phi(y)]$, the tension $T$ is generically not fine tuned.
If we do not impose orbifold boundary conditions in addition to those
above, then there are naively six free parameters: one from the solution on 
each side of the brane of each of equations \eqref{V}-\eqref{A'}.
However, one overall constant shift in $A(y)$ is not relevant because $A(y)$
enters into the equations of motion only through its derivatives.  That leaves
five free parameters and four boundary conditions.  There is generically
a (finite) line of solutions for a region of scalar-brane couplings 
$f[\phi]\,T$.

Once again, this is what we mean by self-tuning: Given an arbitrary
scalar-brane coupling $f[\phi]$ (possibly
satisfying some  constraints), there is a range of ``brane tensions'' $T$
such that static solutions exist.  In the generic case, as argued above, 
there is in fact a continuous set of static solutions for a given boundary
condition specified by $f[\phi]$ and $T$.  Furthermore, if $f[\phi]=f'[\phi]$
then the range of $T$ often extends to infinity.  The reason is that if
$W[\phi]\gg V[\phi]$ asymptotically when $W$ is large, then 
$W'\simeq W$ there, and $T$ can be chosen arbitrarily large such that 
$(fT,f'T)\sim(2W,2W')$.

Orbifold 
boundary conditions are more constraining.  
The additional constraints
from the orbifold condition are $A(y)=A(-y)$ and $\phi(y)=\phi(-y),$ 
which implies that $W_+[\phi]=-W_-[\phi]$, where $W_\pm[\phi]$ are the
solutions for $W$ on the two sides of the brane.
However,
continuity of $A(y)$ and $\phi(y)$ is then guaranteed.  Hence, in this case
there are two free parameters (there would be three but a constant
shift in $A$ is not relevant) and two boundary conditions, and there is
generically a solution for a region of couplings $f[\phi]\,T$.
Thus, there is generically self-tuning in this case, as well.
 
In the absence of orbifold type boundary conditions, if there are $N$ branes
then there are $3N+3-1=3N+2$ free parameters and $4N$ constraints, leaving
$2-N$ free parameters in the solution for a given set of boundary conditions.
Hence, there can be up to two branes without fine-tuning.
With orbifold boundary conditions, where image branes are included in $N$ and
there is assumed to be a brane at the orbifold fixed point (which contributes
1 to $N$), there are $3(N-1)/2+3-1=(3N+1)/2$ free parameters and $4(N-1)/2+
2=2N$ constraints, leaving $(1-N)/2$ free parameters in the solution.  Hence,
only if there is a single brane at the orbifold fixed point will self-tuning
occur. 
If the space is compactified on a circle with orbifold boundary 
conditions,
then a parameter count for the case of branes at the two orbifold fixed points
demonstrates that a fine-tuning is necessary in this case, as well \cite{DZF}. 
Namely,
there are two free parameters in the solution but four boundary conditions.
In what follows we will concentrate on the case of a single 
brane coupled to a scalar field.

%%%%%%%%%%%%%%%%%%%%%%%%%%%%%%%%%%%%%%%%%%%%%%%%%%%%%%%%%%%%%%%%%%%%%%%%
\section{Integrable bulk potentials}
\setcounter{equation}{0}
\setcounter{footnote}{0}

In this section we study some special cases which were also partially 
discussed in
\cite{ADKS,KSS}.  Our discussion of the exact solutions with a vanishing
bulk potential is similar,
but we will not restrict ourselves to the ansatz $A'[\phi]\propto \phi'$
made in \cite{KSS}.
In agreement with the parameter count in the previous section we will
find that there is no fine-tuning in these theories, although some of the
exact solutions exhibit non-generic behavior.

Let us first find the exactly solvable models.  The challenge is finding 
a class of solutions to the nonlinear equation \eqref{V} for a given
$V[\phi]$.
If (in a region where\footnote{The case of a positive bulk potential
can  be studied in a very similar way up to some changes of sign
in the equations \eqref{w1}--\eqref{w2}.} $V<0$)
we write
$W[\phi]$ and $W'[\phi]$ as\footnote{From now, we will denote
by a prime a derivative of $V$, $f$, $W$ or $\omega$ with respect to $\phi$;
or a derivative of $\phi$ or $A$ with respect to $y$.},
\begin{eqnarray}
	\label{w1}
W&=&\sfrac{1}{2} \sqrt{-V[\phi]} \left(w[\phi]+\frac{1}{w[\phi]}\right) ,
\\
	\label{w2}
W'&=&\sfrac{1}{2} \sqrt{-V[\phi]} \left(w[\phi]-\frac{1}{w[\phi]}\right) ,
\end{eqnarray}
then \eqref{V} is immediately satisfied for any {\it prepotential} $w[\phi]$.  The
consistency of \eqref{w1} and \eqref{w2} then translates into a
differential equation for $w[\phi]$:
\begin{equation}
	\label{W'}
\omega'=\omega-\frac{V'}{2V}\, \omega\, \frac{\omega^2+1}{\omega^2-1}
\end{equation}
Exact solutions for $w[\phi]$
can be found when \eqref{W'} is separable, {\em i.e.}
when $V'[\phi]/V[\phi]$ is a constant.

We distinguish three cases:
\begin{itemize}
\item a vanishing bulk potential: $V[\phi]=0.$
\item a negative bulk cosmological constant: $V[\phi]=\Lambda <0.$
\item an exponential bulk potential: $V'[\phi]/V[\phi]= {\rm const.} \neq 0.$
\end{itemize}

%%%%%%%%%%%%
\subsection{Vanishing bulk potential}

This case has been extensively studied by refs. \cite{ADKS,KSS}
but it is a worthwhile exercise to repeat the discussion
in terms of a superpotential.
The equation for the superpotential can be solved without introducing
a prepotential. Indeed, eq. \eqref{V} becomes simply
\begin{equation}
	\label{eq:V=0W}
W'[\phi]^2= W[\phi]^2,
\end{equation}
with two branches of solutions,
\begin{equation}
W[\phi]=c\, e^{\epsilon\phi}.
\end{equation}
where $c$ is a constant of integration and $\epsilon$ is a sign, both of them
can take different values on the two sides of the brane (the constants of integration
relative to the right (left) hand side of the brane will be denoted
with a $+$ ($-$) subscript).
With this form of the superpotential, the eqs. \eqref{phi'}-\eqref{A'} can
be easily solved as
\begin{eqnarray}
	\label{eq:V=0phi}
&&
\phi (y) = - \epsilon \ln (d-cy) \\
	\label{eq:V=0A}
&&
A (y) = -\ln (d-cy) + e
\end{eqnarray}
where $d$ and $e$ are some constants of integration that can also differ
on the sides of the branes. Moreover, to make sense eq. \eqref{eq:V=0phi} need:
$d_+>0$ and $d_->0$.
Thus the continuity requires
\begin{eqnarray}
	\label{eq:V=0jump1}
&&
\phi(0)\equiv \phi_0 = - \epsilon_+ \ln d_+ = - \epsilon_- \ln d_-
\\
	\label{eq:V=0jump2}
&&
A(0)\equiv A_0 = e_+ - \ln d_+ =  e_- - \ln d_-
\end{eqnarray}
while the jump equations are
\begin{eqnarray}
&&
\frac{\epsilon_+ c_+}{d_+} - \frac{\epsilon_- c_-}{d_-} = f'[\phi_0]\, T
\\
&&
\frac{c_+}{d_+} - \frac{c_-}{d_-} = f[\phi_0]\, T
\end{eqnarray}
From the expression of the warp factor, we conclude that
the Planck scale on the brane,
$\kappa^{-2}_{D-1}=\kappa^{-2}_{D}\int dy\, e^{-(D-3)A/(D-1)}$,
is finite {\it iff} singularities are encountered on both sides
of the brane {\it i.e.} $c_+>0$ and $c_-<0$.
\begin{itemize}
\item if $\epsilon_+ \epsilon_-=1$:~
the consistency of the jump equations requires that $f'[\phi_0]=\epsilon f[\phi_0]$
and then it is possible to find solutions with or without singularities
for any value of the brane tension but the singular
solutions correspond to $f[\phi_0]\, T >0$.
\item if $\epsilon_+\epsilon_-=-1$:~
the solutions with singularities exist only if
$f[\phi_0]\, T >0$ and $-1<f'[\phi_0]/ f[\phi_0]<1$ but do not require any
fine-tuning.
\end{itemize}

If we choose $f[\phi]=Ce^{\epsilon\phi}$ for the case $\epsilon=\epsilon_+=\epsilon_-$
then it is clear that
the boundary conditions can be satisfied for any $T$. There are several
important comments to make about this case, which is quite non-generic.  
First of all, the fact that
a specific form had to be chosen for $f[\phi]$ is a result of two non-generic
features of the model:  First, there is a
symmetry $\phi\rightarrow \phi+\mbox{\it const.}$.  As
a result of this symmetry, one of the free parameters, namely $\phi_0$,
does not appear on the left hand side of the boundary conditions for
the derivatives $\phi'$ and $A'$.  In addition, it turned out that
the left hand sides of the two boundary conditions had the same
form up to a sign.  As a result, only $f'[\phi]/f[\phi]$ is relevant, and
given one solution $(f[\phi] T,f'[\phi]T)$, there is an infinite set of 
solutions with the {\em same} $f[\phi]$ and arbitrary $T$.  The fact that 
$T$ is completely arbitrary is most likely unique to the case of vanishing 
bulk potential.  But the fact that given $f,f'\sim{\cal O}(M_5)$, self-tuning
occurs for $T$ to within 
${\cal O}(M_{5})\gg {\cal O}(M_{EW})$ (where ${\cal O}(M_{EW})$ is the
expected Standard Model contribution to the tension) is generic, and is what 
we mean by  self-tuning.

Even given the constraint on $f[\phi]$ from the shift symmetry in $\phi$
in this case, as explained in \cite{ADKS,KSS} the required
exponential form  of $f[\phi]$ might be natural from a stringy perspective
where $\phi$ is interpreted as the dilaton.

Furthermore, as pointed out in \cite{KSS}, the solutions with 
singularities on either side of the brane must be chosen in order to have
a finite gravitational coupling (assuming that the spacetime can be cutoff
at the singularity in a consistent way).  This feature will
turn out to be generic except in theories which admit solitonic solutions
for fine-tuned boundary conditions {\it i.e.} for special discrete values
of the brane tension.

%%%%%%%%%%%%
\subsection{Bulk cosmological constant}

Even though the case of a (negative) bulk cosmological constant
can be studied without introducing a prepotential, we will
present our method on this rather simple example before
proceeding to the somewhat more complicated case
of an exponential bulk potential in the next subsection.
We will denote by $V[\phi]=\Lambda<0$ the cosmological constant.
The equations of motion are:
\begin{eqnarray}
\frac{d\omega}{d\phi} & = & \omega \ ,
\\
\frac{d\phi}{dy} & = & \sfrac{1}{2}\, \sqrt{-\Lambda}\, \omega (1-\omega^{-2})\ ,
\\
\frac{dA}{dy} & = & \sfrac{1}{2}\, \sqrt{-\Lambda}\, \omega (1+\omega^{-2}) \ .
\end{eqnarray}
The differential equation for the prepotential can be easily integrated,
\begin{equation}
\omega = c\, e^\phi \ ,
\end{equation}
where $c$ is a constant of integration. Plugging this expression
for the prepotential into the remaining equations of motion, we obtain,
\begin{eqnarray}
\phi (y) & = & -\ln ( c\, \vartheta (y)) \ ,
\\
A (y) & = & - \ln |\vartheta (y)| + \ln |1-\vartheta^2 (y)| + a \ ,
\end{eqnarray}
where $a$ is a constant of integration and another constant of integration, $y_c$,
also appears in the expression of the function $\vartheta(y)$ defined, on the
two different branches of solutions, by
\begin{equation}
\vartheta (y) = - \tanh \sfrac{\sqrt{-\Lambda}}{2} (y-y_c)
\ \ \ \mbox{or } \ \ \
\vartheta (y) = - \coth \sfrac{\sqrt{-\Lambda}}{2} (y-y_c) \ .
\end{equation}
The nature of the solution depends on the sign of $y_c$ on each side
of the brane. On the right (left) hand side, a positive (negative)
value of $y_c$ will correspond to a solution involving a singularity
at a finite proper distance from the brane, $y=y_c$. Near this singularity,
the warp factor, $\exp(-2A/(D-1))$, goes to zero so the singularity appears
as an horizon in the bulk. Conversely, if $y_c^+$, the value of $y_c$ on the
right hand side of the brane, is negative (respectively, $y_c^->0$),
then the transverse dimension will be infinitely large; however, near infinity,
the warp factor blows up and the Planck scale on the brane
diverges, ruining any phenomenological relevance.

The values of the constants of integration are constrained
by the continuity and jump equations
(with $\vartheta_\pm= \tanh \sqrt{-\Lambda}y_c^\pm/2$),
\begin{eqnarray}
&&
\phi_0 \equiv -\ln (c_+\vartheta_+) = -\ln (c_-\vartheta_-)\ ,
\\
&&
A_0 \equiv - \ln |\vartheta_+| + \ln |1-\vartheta^2_+| + a_+
= - \ln |\vartheta_-| + \ln |1-\vartheta^2_-| + a_- \ ,
\\
&&
\sfrac{1}{2}\,\sqrt{-\Lambda}\,
\left( \sfrac{1+\vartheta_+^2}{\vartheta_+} -
\sfrac{1+\vartheta_-^2}{\vartheta_-} \right) = f[\phi_0]\, T  \ ,
\\
&&
\sfrac{1}{2}\,\sqrt{-\Lambda}\,
\left( \sfrac{1-\vartheta_+^2}{\vartheta_+} -
\sfrac{1-\vartheta_-^2}{\vartheta_-} \right) = f'[\phi_0]\, T \ .
\end{eqnarray}
A solution to these equations can be found, provided that
$f'[\phi_0]\not = \pm f[\phi_0]$, for any value of the brane tension such that
\begin{equation}
T^2 > \frac{-4\Lambda}{f[\phi_0]^2-f'[\phi_0]^2}.
\end{equation}
Moreover, singularities will exist on both sides of the brane
{\it iff}
\begin{equation}
f[\phi_0]\, T >0 \ \ \ \mbox{and } \ \ \ -1< \frac{f'[\phi_0]}{f[\phi_0]}<1 ,
\end{equation}
just as in the case of a vanishing bulk potential.

It is interesting to look at the $\mathbb{Z}_2$ symmetric solution for which
there are the additional constraints:
\begin{equation}
y_c^+ = -y_c^-, \ \ \ c_+ = - c_- \ \ \ \mbox{and } \ \ \ a_+ = a_- .
\end{equation}
The continuity conditions are automatically satisfied but the consistency
of the jump equations requires a fine-tuning,
\begin{equation}
	\label{Z2finetuned}
f[\phi_0]^2-f'[\phi_0]^2 = -\frac{4 \Lambda}{T^2} .
\end{equation}
As already discussed in the case with a vanishing bulk potential,
this fine-tuning is a consequence of the translational symmetry,
$\phi\rightarrow \phi+ \mbox{\it const.}$, in the theory.
As before, because of this
shift symmetry we
lose the appearance of a free parameter to adjust
in the jump equations, which leads to a
more restricted set of boundary conditions than for a generic
bulk potential $V[\phi]$. The fine-tuning \eqref{Z2finetuned}
is precisely the one appearing in the Randall--Sundrum model
when the scalar coupling to the brane is a constant:
\begin{equation}
	\label{eq:LbkLbr}
\Lambda_{bk} = -\frac{D-1}{8(D-2)} \kappa_D^2 T_{br}^2,
\end{equation}
where $\Lambda_{bk}$ and $T_{br}$ are the
canonically normalized quantities \cite{CGS3},
\begin{equation}
\Lambda_{bk} = \frac{D-2}{2(D-1)\kappa_D^2} \, \Lambda
\ \ \ \mbox{and } \ \ \
T_{br} = \frac{D-2}{(D-1)\kappa_D^2} \, f[\phi_0] T .
\end{equation}

The solution constructed by Randall and Sundrum that localizes gravity
with an infinitely large extra dimension corresponds to a limit
of the singular solution where the singularities are pushed to infinity
{\it i.e.} $\vartheta_\pm=\pm 1$. The jump equations then require
\begin{equation}
f'[\phi_0]=0 \ \ \ \mbox{and } \ \ \ f[\phi_0]\, T = 2 \sqrt{-\Lambda}
\end{equation}
{\it i.e.} the coupling between the brane and the scalar field vanishes
and the canonically normalized brane tension, $T_{br}$, is fine-tuned
to \eqref{eq:LbkLbr}. In this limit, the expressions for the scalar
field and the warp factor simply become
\begin{equation}
\phi=\phi_0 \ \ \ \mbox{and } \ \ \
A = \sqrt{-\Lambda}\, |y| .
\end{equation}
%

%%%%%%%%%%%%
\subsection{Exponential bulk potential}
\label{sec:exp}
The last case that can be solved analytically with our method
involves an exponential potential for the scalar field in the bulk.
We will concentrate on negative potential while the case of positive
potential requires some minimal changes. So the bulk potential will be 
parametrized by two real numbers $a$ and $b$:
\begin{equation}
 V[\phi] = -a^2 e^{2b\phi} \ .
\end{equation}

On each side of the brane, the equations of motion are simply:
\begin{eqnarray}
\frac{d\omega}{d\phi} & = & \omega-b\,\omega\,\frac{\omega^2+1}{\omega^2-1} \ ,
\\
\frac{d\phi}{dy} & = & \sfrac{1}{2}\, a e^{b\phi}\, \omega (1-\omega^{-2}) \  ,
\\
\frac{dA}{dy} & = & \sfrac{1}{2}\, a e^{b\phi}\, \omega (1+\omega^{-2}) \ .
\end{eqnarray}
The sign of $a$ is not fixed and can be chosen independently on the two
sides of the brane, as we will discuss.
The first differential equation can be easily solved to express $\phi$
in terms of $\omega$:
\begin{equation}
e^{-b\phi} = e^{-bc}\, {\left| \omega \right|}^{-b/(1+b)}\,
{\left| 1+b-(1-b)\omega^2 \right|}^{b^2/(b^2-1)} \ ,
\end{equation}
where $c$ is a constant of integration. This last result can be used to obtain
a parametric representation of $A$ and $y$ as functions of $\omega$:
\begin{eqnarray}
	\label{eq:Yw}
&&
y(\omega)  =
-\sfrac{2}{a} e^{-bc}\,
\int_{\omega_0}^{\omega} d\tomega \
\frac{
{\left| \tomega \right|}^{-b/(1+b)}\,
{\left| 1+b-(1-b)\tomega^2 \right|}^{b^2/(b^2-1)}}
{{\left( 1+b-(1-b)\tomega^2 \right)}} \ ,
\\
	\label{eq:Aw}
&&
A(\omega)  =  A_0
- \int_{\omega_0}^{\omega} d\tomega\
\frac{(1+\tomega^2)}
{\tomega\,
{\left( 1+b-(1-b)\tomega^2 \right)}} \ .
\end{eqnarray}
On the two sides of the brane, the parameter $a$ can differ by a sign while
the initial bound of integration, $\omega_0$, and the constant of integration,
$c$, can take any values compatible with the continuity conditions.
A $\mathbb{Z}_2$ symmetric solution will correspond to two different choices
of sign for $a$ but the same values for $\omega_0$ and $c$.

Different kinds of solutions can be obtained depending on the value
of $b$ and of the range of integration for the variable $\tomega$:
\begin{itemize}
\item $b<-1$~: in that case, $dy/d\omega$ has an integrable singularity at
$\omega=+\infty$ but a non-integrable singularity at $\omega=0$  while
the singularities of $dA/d\omega$ are both non-integrable. So we can find
solution with or without bulk singularity at finite proper distance.
\begin{itemize}
\item Solutions without singularity will be given by (\ref{eq:Yw})-(\ref{eq:Aw})
where on the right (left) hand side of the brane, $a$ has to be chosen positive
(negative) and $\omega_0$ is a negative initial bound of integration.
The parameter $\omega$ will range from $\omega_0$ to $0^-$.
It is easy to find the asymptotic behavior of this solution for large
$|y|$, {\it i.e.} $\omega$ near $0^-$:
\begin{equation}
A \underset{|y|\sim \infty}{\sim} -\ln |y| \ ,
\end{equation}
from where it becomes evident that the singularity free solutions do not
localize gravity because the Planck scale on the brane diverges.
\item Solutions with singularity will still be given by eqs.
(\ref{eq:Yw})-(\ref{eq:Aw}) with a positive (negative) parameter $a$
on the right (left) hand side of the brane. And the range of integration
goes from a positive initial value, $\omega_0$,
to $+\infty$. In that case, $y$ reaches a finite value $y_c$ while
the warp factor goes like:
\begin{equation}
A \underset{y\sim y_c}{\sim} -\ln |y-y_c| \ ,
\end{equation}
which indicates that the singularity is a horizon where the metric on the brane
vanishes. The behavior of the warp factor near the singularity insures
that the Planck scale on the brane,
$\kappa^{-2}_{D-1}=\kappa^{-2}_{D}\int dy\, e^{-(D-3)A/(D-1)}$,
is finite.
\end{itemize}
\item $-1<b<1$~: in that case the singularities of $dy/d\omega$ at
$\omega=\pm\infty$, $\omega=\pm\sqrt{(1+b)/(1-b)}$ and $\omega=0$
are integrable while those appearing in $dA/d\omega$ are non-integrable.
All the solution are singular with a horizon or a curvature singularity.
\item $1<b$~: this case is quite similar to the first case because
the singularity of $dy/d\omega$ at $\omega=0$ is integrable
but the singularity at $\omega=+\infty$ is non-integrable while
the two singularities of $dA/d\omega$ are both  non-integrable.
So we can construct solutions with or without singularity.
\begin{itemize}
\item Solutions without singularity will be given by (\ref{eq:Yw})-(\ref{eq:Aw})
where on the right (left) hand side of the brane, $a$ has to be chosen negative
(positive) and $\omega_0$ is a positive initial bound of integration.
The parameter $\omega$ will range from $\omega_0$ to $+\infty$.
It is easy to find the asymptotic behavior of this solution for large
$|y|$, {\it i.e.} $\omega$ near $+\infty$:
\begin{equation}
A \underset{|y|\sim \infty}{\sim} -\ln |y| \ ,
\end{equation}
from where, once again, it becomes evident that the singularity free solution does
not localize gravity.
\item Solutions with singularity will still be given by eqs.
(\ref{eq:Yw})-(\ref{eq:Aw}) with  a negative (positive) parameter $a$
on the right (left) hand side of the brane. And the range of integration
goes from a negative initial value, $\omega_0$,
to $0^-$. In that case, $y$ reaches a finite value $y_c$ while
the warp factor goes like:
\begin{equation}
A \underset{y\sim y_c}{\sim} -\ln |y-y_c| \ ,
\end{equation}
which indicates that the singularity is a horizon where the metric on the brane
vanishes.
\end{itemize}
\item $b=\pm1$~: in these two cases, we can construct singularity free
solutions with a blowing up warp factor as well as singular solutions
with a horizon at a finite proper distance.
\end{itemize}

\begin{figure}[htb]
\centerline{\epsfxsize=15cm\epsfbox{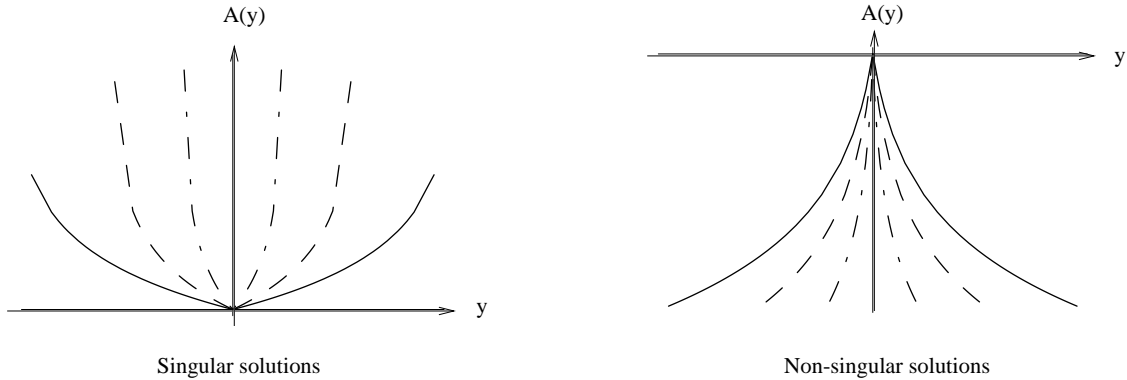}}
\caption{ Shapes of the warp factor for singular and non-singular solutions
to the equations of motion with an exponential bulk potential.
We have drawn $\mathbb{Z}_2$ symmetric
solutions for different values of the tension on the brane: when
the tension increases, the horizon becomes closer and closer to the brane
for the singular solutions while the warp factor, $e^{-2A/(D-1)}$, blows up
faster and faster for the the singularity-free solutions.
In both cases, the jump conditions require that the scalar coupling to the brane
satisfies: $|(df/d\phi)/f|<1$ on the brane.
}
\label{fig:cc-exp}
\end{figure}

Figure \ref{fig:cc-exp} illustrates the different types of solutions.

\vspace{.2cm}

The main result of the analysis of these integrable bulk potentials
is that we find two kinds of solutions:
{\it (i)} solutions with a horizon in the bulk at a finite proper distance
from the brane  and with a finite lower dimensional Planck scale;
{\it (ii)} solutions without singularity at a finite proper distance
but associated to a bulk geometry that decouples gravity on the brane.
Both kinds of solutions do not require any fine-tuning and can be constructed
for a range of brane tension, $T$. However it seems impossible to find a singularity
free solution that localizes gravity on the brane, unless the brane tension is
fine-tuned as in the Randall--Sundrum model. This point surely deserves
further scrutiny which the next sections will be devoted to.

%%%%%%%%%%%%%%%%%%%%%%%%%%%%%%%%%%%%%%%%%%%%%%%%%%%%%%%%%%%%%%%%%%%%%%%%%%%%%%
\section{Perturbative and numerical methods}
\setcounter{equation}{0}
\setcounter{footnote}{0}

In this section we give two methods for finding approximate solutions 
to the coupled equations of motion in the bulk that satisfy the 
boundary conditions at the brane. First, we present a perturbative
method, which we then show to break down for the interesting case 
of perturbing around a solution with localized gravity. Then we give a 
systematic numerical method for solving the equations.

%%%%%%%%%%%%%%%%%%%%%%%%%%%%%%%%%%
\subsection{A perturbative method}

As emphasized in the previous section, the key to finding the self-tuned
solutions is to utilize the fact that one can transform the superpotential
without changing the bulk potential for the scalar field. Thus, one has 
an additional degree of freedom if one picks a different superpotential 
function on the two sides of the brane, $W_+[\phi]$ and $W_- [\phi]$.
Let us now assume that we have found a static solution to the equations of 
motion $\phi_0 (y)$ and $A_0(y)$, and assume that we have chosen one of the
integration constants such that $A_0(0)=0$. This is a solution for a 
particular value $T$ of the brane tension. In order to find the solution
obtained by perturbing the brane tension as $T\to T +\delta T$, we first need
to find how one can change the superpotential $W$ around $W_0 [\phi]$
so as to leave the bulk potential unchanged:
\begin{equation}
V[\phi]=W_0'[\phi]^2-W_0[\phi]^2,
\label{potential}
\end{equation}
which should be invariant under $W\to W+\delta W$.
Linearizing (\ref{potential}) around $W_0$ we obtain a differential 
equation for $\delta W$ of the form
\begin{equation}
\frac{\delta W'}{\delta W}= \frac{W_0}{W_0'},
\end{equation}
which is solved by
\begin{equation}
	\label{pert}
\delta W [\phi] = C\, \exp \int \frac{W_0 [\phi]}{W_0' [\phi]} d\phi,
\end{equation}
where the arbitrary constant $C$ yields the extra degree of freedom
needed to find a solution for any value of $T$. The superpotential variation
can be expressed in terms of the unperturbed background solution using
the equations of motion $d\phi_0/W_0'[\phi_0]=dy$ and
$W_0 [\phi_0]=dA_0/dy$, we obtain
\begin{equation}
	\label{perturb}
\delta W [\phi (y)]= C\, \exp  \int dy \frac{dA_0}{dy}
= C\, e^{A_0 (y)}.
\end{equation}
From (\ref{pert}) we can see that the change in the derivative of the
superpotential is given by
\begin{equation}
\delta W' [\phi (y)] = C\, \frac{A_0' (y)}{\phi_0' (y)}\, e^{A_0 (y)}.
\end{equation}
In order to satisfy the jump equations \eqref{bc} for the perturbed tension
$T+\delta T$, we need to choose $C$ differently on the two sides of the
brane. The values of $C_{\pm}$ are then given by
\begin{equation}
	\label{Cpm}
C_{\pm}=\frac{f'[\phi_0] - f[\phi_0]\frac{A'_{0\mp}(0)}{\phi'_{0\mp }(0)}}
{\Delta \frac{A'_0}{\phi'_0}} \ \delta T,
\end{equation}
where $\Delta\, A'_0/\phi'_0$ denotes the jump in $A'_0/\phi'_0$
at $y=0$. Once $C_{\pm}$ is
determined from (\ref{Cpm}), we can simply integrate the equations
\begin{equation}
	\label{pertA}
\delta \phi' (y) = \delta W' [\phi_0 (y)], \ \
\delta A' (y)= \delta W [\phi_0 (y)]
\end{equation} 
to obtain the perturbed solutions $\phi_0 (y)+\delta\phi (y)$ and
$A_0(y)+\delta A(y)$. This method
always results in a perturbed solution. However,
in the most interesting case, when the unperturbed solution
asymptotes to AdS space thereby localizing gravity to the brane (that is
for $A(y) \sim (D-1)|y|/R_{AdS}$ for large values of $y$), one can easily
see that the perturbative method presented here always breaks down.
This can be seen by inspecting (\ref{perturb}), which shows that
in this case $\delta W [\phi (y)] \propto e^{(D-1)|y|/R_{AdS}}$. Therefore the
perturbed values of $\delta \phi$ and $\delta A$ grow exponentially,
and thus the linearized approximation breaks down. We will examine
the case of localized gravity in detail in Section \ref{sec:no-go}. But first we
give a numerical method that can be used for any choice of the bulk potential.
%%%%%%%%%%%%%%%%%%%%%%%%%%%%%%%

%%%%%%%%%%%%%%%%%%%%%%%%%%%%%%
\subsection{A numerical method}

Next we present a method for solving the system (\ref{V})-(\ref{A'}) 
numerically for any brane tension and potential in the bulk. 
Thus the input functions are the bulk potential $V[\phi]$, the brane
tension $T$, the coupling of the scalar to the brane determined by
the function $f[\phi]$, and in addition we can pick the value of the
scalar field at the brane $\phi (0)\equiv \phi_0$ arbitrarily.
In order to find a numerical 
solution to these equations, one has to first make sure that the boundary
conditions that one imposes do satisfy the jump equations (\ref{bc}). 
Our strategy is the following: we first determine the superpotential
functions $W_+[\phi]$ and $W_-[\phi]$ to the left and the right of the
brane numerically such that the boundary conditions arising from
the coupling to the brane are satisfied. This can be done by noting that
once $\phi_0$ is fixed, the jump equations are just given by
\begin{equation}
	\label{jump}
W_+-W_-=f_0 T, \ \ \ \ W_+'-W_-'=f_0' T,
\end{equation}
where $W_{\pm}$ refers to the values of the superpotential functions
to the right and left of the brane at $\phi_0$, $f_0=f[\phi_0]$, etc.
In addition, the superpotential functions must be such that they reproduce 
the correct value of the bulk potential at the brane:
\begin{equation}
	\label{value}
{W_+'}^2-W_+^2=V_0, \ \ \ \ {W_-'}^2-W_-^2=V_0,
\end{equation}
where $V_0=V[\phi_0]$. Eqs. (\ref{jump}) and (\ref{value}) together are
enough to determine the values of both $W_{\pm}$ and $W_{\pm}'$ at the branes.
They are given by the expressions:
\begin{eqnarray}
	\label{numsol}
&& W_{\pm}=\pm \frac{1}{2} f_0 T +\frac{1}{2} f_0' \sqrt{T^2+\frac{4 V_0}{
{f_0'}^2-f_0^2}} \nonumber \\
&& W_{\pm}'=\pm \frac{1}{2} f_0' T +\frac{1}{2} f_0 \sqrt{T^2+\frac{4 V_0}{
{f_0'}^2-f_0^2}}.
\end{eqnarray}
Due to the quadratic nature of equations (\ref{jump}) and (\ref{value})
there is a second solution, where the signs in front of the 
square roots in (\ref{numsol}) are both simultaneously flipped. 
Once the value of $W_{\pm}$ and $W'_{\pm}$ are fixed, one can numerically
integrate the equation\footnote{We will see in Section 5 that, unless the brane
tension is fine-tuned, the superpotential, $W[\phi]$, will be a monotonic
function of $\phi$ and thus $W'_\pm[\phi]$ will keep the sign of $W'_\pm$.}
\begin{equation}
W_{\pm}' [\phi] = {\rm sgn} (W_{\pm}') \sqrt{V[\phi]+W_{\pm}[\phi]^2}
\end{equation}
to obtain the superpotential functions to the left and the right of the
brane that satisfy all boundary conditions. Once $W_{\pm} [\phi]$
are numerically known, we can simply integrate the equations
\begin{equation}
\phi' (y) = W' [\phi (y)], \ \ A' (y)=  W [\phi (y)]
\end{equation} 
to the left and the right of the brane to obtain the numerical
solutions for $\phi (y)$ and $A(y)$. We will show an example for this
below for the case when the exact solution is known, and compare the two 
results. 

\subsection{An example for the perturbative method}
In this Section we test how well the perturbative method described 
in 4.1 works. We will compare the analytic solution of the model with a 
vanishing bulk potential to the perturbed solution around a different
analytic solution. The main conclusions are as expected:
the perturbative method works well far from the singularities. However it
gets worse as we approach the singularity itself, and does not capture
the essential feature of the self-tuning solutions: whereas in the self-tuning
mechanism, for a fixed value of the scalar field on the brane, the place of the
singularity adjusts itself with respect to the value of the brane tension, here
the singularity of the
perturbed solution remains at the same place where the singularity of the
unperturbed solution was. Therefore, we conclude that this method is not
very efficient in capturing the basic properties of the self-tuning
solutions.

\begin{figure}[htb]
\centerline{\epsfxsize=7cm\epsfbox{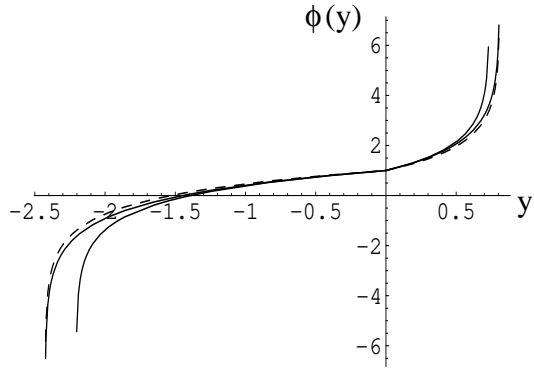}}
\caption{The exact solution versus the perturbed solution for the case of a
vanishing bulk potential. The curve that blows up at smaller values of 
$y$ corresponds to the exact solution. The initial unperturbed solution
is given by the dashed curve.
The singularity of the
perturbed solution appears at the same distance from the brane as the 
singularity
of the initial solution; the shift of the singularity with a variation
of the tension is missing. }
\label{fig:perturbed}
\end{figure}

The example we consider is the vanishing bulk potential discussed in
3.1, with the choice $\epsilon_+=1, \epsilon_-=-1$, and we choose 
$f[\phi]=e^{\phi /2}$, $\phi_0=1$, and for the unperturbed solution
we choose $T=1$, while we pick $\delta T =0.1$ for the perturbed solution.
The analytic solution is given by \eqref{eq:V=0phi}--\eqref{eq:V=0A}, with
\begin{equation}
d_{\pm} =e^{\mp \phi_0}, \ \ c_+=\frac{3}{4} T e^{-\phi_0/2}, \ \ 
c_-=-\frac{1}{4} T e^{ 3 \phi_0/2}, \ \ e_\pm =\mp \phi_0,
\end{equation}
where $A$ has been normalized to zero on the brane.
The perturbed solution from \eqref{pertA} is given by
\begin{equation}
\delta \phi_+ (y) = -\frac{\delta T}{T} \ln \left(\frac{d_+-c_+ y}{d_+}\right),
\ \ \
\delta \phi_- (y) =\frac{\delta T}{T} \ln \left(\frac{d_--c_- y}{d_-} \right).
\end{equation}
where $\delta \phi_\pm$ has been normalized to zero on the brane.
The perturbed solution obtained for $T=1, \delta T=0.1$ compared to the
exact solution for $T=1.1$ can be seen in Fig. \ref{fig:perturbed}. As
mentioned above, the perturbative solution nicely follows the exact 
solution away from the singularity, but deviates from it close to the
singularity, in particular the place of the singularity is incorrectly
predicted to coincide with the singularity of the unperturbed solution.

%%%%%%%%%%%%%%%%%%%%%%%%%%%%%%%%%%%%%%%%%%
\subsection{An example for the numerical method}

We have shown above, that the perturbative method in general does not
do a good job in finding the solutions, since it becomes unreliable close to
the singularities. However, the numerical solution should not have these 
problems. Indeed, we analyze the same example as above 
(the case with vanishing bulk potential) using the 
numerical method, and find that the exact and numerical curves are 
virtually indistinguishable. Therefore, we suggest that in order to
analyze potentials for which no exact solutions can be found, one should
use the numerical method rather than the method based on perturbations.

\begin{figure}[htb]
\centerline{\epsfxsize=7cm\epsfbox{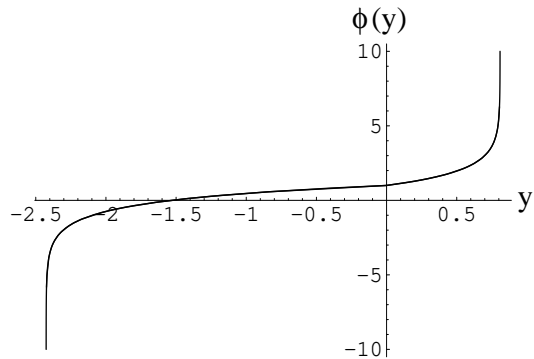}}
\caption{The numerical solution overlayed on the exact solution for 
the case of a vanishing bulk potential. The fact that the two curves are
indistinguishable shows that the numerical method works extremely well even
close to the singularities.}
\label{fig:numeric}
\end{figure}

We are looking for a numerical solution to the case analyzed perturbatively
above, that is vanishing bulk potential, $f[\phi]=e^{\phi/2}$, $\phi_0 =1$,
$T=1$ and $\epsilon_+=1, \epsilon_-=-1$.
From Eqs. \eqref{eq:V=0W} and \eqref{eq:V=0jump1}
we find the starting values of the superpotential
to the left and the right of the brane:
\begin{equation}
W_+= \frac{3}{4} e^{\phi_0 /2} T, \ \ W_-= -\frac{1}{4} e^{\phi_0 /2} T.
\end{equation}
Numerically integrating the equation
\begin{equation}
W_\pm'[\phi] = \pm W[\phi]
\end{equation}
with the boundary conditions $W_\pm [\phi_0]=W_\pm$ one obtains the
numerical values for $W_\pm (\phi)$. Finally, the values for $\phi (y)$ 
can be obtained by numerically inverting the integral
\begin{equation}
y=\int_{\phi_0}^{\phi} \frac{d\phi }{W' (\phi )}
\end{equation}
to the left and right of the brane.
The numerical solution obtained this way overlayed on the exact solution
of Section 2 can be seen in Fig. \ref{fig:numeric}. One can see that the
two curves are virtually indistinguishable, suggesting that the numerical
method works very well around the singularities, and should be the 
preferred method of looking for solution in the absence of exact solutions.

%%%%%%%%%%%%%%%%%%%%%%%%%%%%%%%%%%%%%%%%%%%%%%%%%%%%%%%%%%%%%%%%%%%%%%%%%%%%%%%%%%
\section{Localized gravity without singularities}
\label{sec:no-go}
\setcounter{equation}{0}
\setcounter{footnote}{0}
%%%%%%%%%%%

\subsection{A no-go theorem}
We would like to reexamine in this section the count of free parameters versus
fine-tuning parameters needed to preserve a static Poincar\'e invariance on 
the brane,
with the restriction that the solution corresponds to an infinitely large
extra-dimension (without singularities) in the bulk and localizes gravity on 
the brane.

Consider the general D-dimensional background preserving
$\mbox{\it Poincar\'e}_{D-1}$
\begin{equation}
ds^2 = e^{-2 A(y)/(D-1)} dx^2_{D-1} + dy^2 \ .
\end{equation}
The graviton zero-mode is localized on the brane\footnote{We do not consider 
the recently proposed possibility that gravity might be quasi-localized to the
brane\protect\cite{quasilocal}, since in those models the $A''>0$ condition
is not satisfied, therefore it is not possible to generate those backgrounds
from a single scalar field.} precisely when the effective 
Planck scale is
finite on the brane \cite{CEHS}. In terms of the warp factor, this condition 
is:
\begin{equation}
	\label{MPl}
\frac{1}{\kappa_{D-1}^2} =
\frac{1}{\kappa_{D}^2}
\int dy\ e^{-(D-3)A(y)/(D-1)}
\ < \ \infty  \ .
\end{equation}
It is convenient to introduce a transverse coordinate $z$ for which the bulk metric
is conformally flat:
\begin{equation}
ds^2 = \Omega^2(z) \left( dx^2_{D-1} + dz^2 \right) \ ,
\end{equation}
where the conformal factor, $\Omega$, is related to the warp factor,
$e^{-2 A\left(y\left(z\right)\right)/(D-1)}$, by the two identities:
\begin{equation}
\Omega (z) = e^{- A(y)/(D-1)}
\ \ \ \mbox{and } \ \
\Omega^2 (z) dz^2 = dy^2  \ .
\end{equation}
Then the condition (\ref{MPl}) is equivalent to having a massless normalizable
bound state, which is interpreted as the graviton on the brane:
\begin{equation}
	\label{graviton}
\psi_0\propto \Omega^{(D-2)/2}
\ \ \ \mbox{with } \ \ \
\int dz\ |\psi_0|^2 < \infty \ .
\end{equation}
Let us assume that the behavior of $\psi_0$ at infinity is a power law:
\begin{equation}
\psi_0\underset{z\sim \infty}{\propto} z^{-\alpha} \ .
\end{equation}
The localization of gravity (\ref{graviton}) then requires:
$\alpha>1/2$.

In our study, the value of the parameter $\alpha$ is
constrained by the fact
that the background is created by a scalar field coupled to gravity. From the
equations of motion, we easily deduce that $A$ has to satisfy:
$d^2 A/dy^2 \geq 0$, which translates, in the $z$ coordinate, in a lower bound
on the value of $\alpha$:
\begin{equation}
\alpha \geq \frac{D-2}{2} \ .
\end{equation}

Furthermore it is worth noticing that an upper bound on $\alpha$ comes
by the requirement of a geometry without singularity at a finite
proper distance\footnote{For a power law conformal factor, the curvature
always vanishes at infinity. However quadratic invariants such as
$R_{M_1M_2M_3M_4}R^{M_1M_2M_3M_4}$ will be singular at infinity
as soon as $\alpha>(D-2)/2$.}.
Indeed the proper distance from the brane to infinity is given:
$l_\infty= \int dz \,\Omega$
that diverges {\it iff}:
\begin{equation}
\alpha \leq \frac{D-2}{2} \ .
\end{equation}

So the only background for the scalar field coupled to gravity that localizes
gravity without singularity (with an infinitely large extra dimension)
is asymptotic, at infinity, to the horizon of an anti-de Sitter space, as in
the RS model, and corresponds
to $\alpha=(D-2)/2$. In that case, the warp factor is exponentially decreasing
with the proper distance to the brane:
\begin{equation}
A \underset{|y|\sim \infty}{\sim} (D-1) |y| / R_{AdS} \ .
\end{equation}

The aim of this section is to show that such a background necessarily requires
a fine-tuning between the brane and the bulk.  This is not to say that there
is no self-tuning in these models, only that the nonsingular solutions require
fine-tuning.

The previous asymptotic behavior has a nice interpretation in terms
of the superpotential, $W[\phi]$. According to the equations of motion,
the fact that $A$ is asymptotically linear means that $\phi$ becomes
constant and we will denote by $\phi^-_c$ and $\phi^+_c$ the asymptotic
values of $\phi$ at $y=-\infty$ and $y=+\infty$ respectively.

The equation
\begin{equation}
	\label{eq:phi}
\frac{\partial \phi}{\partial y} = \frac{dW}{d\phi} ,
\end{equation}
is similar to an RGE with $W'$ playing the role of the $\beta$-function.
In order for $\phi$ to 
approach a constant (fixed point)
at infinity, the $\beta$-function $dW/d\phi$ must have zeroes.
In other words, in order to extend the range of the transverse coordinate from
$y=-\infty$ to $y=+\infty$, the values $\phi^-_c$ and $\phi^+_c$ at infinity
must be some roots of $dW/d\phi$:
\begin{equation}
{\frac{dW}{d\phi}}\Big|_{\phi^-_c} =0
\ \ \ \mbox{and } \ \ \
{\frac{dW}{d\phi}}\Big|_{\phi^+_c} =0 \ .
\end{equation}
Furthermore, at infinity, $A$ has to be linearly {\it increasing} in $|y|$;
otherwise the conformal infinity of AdS would be reached without localized
gravity \cite{CGS3}. Given that
\begin{equation}
	\label{eq:A}
\frac{\partial A}{\partial y} = W[\phi] ,
\end{equation}
we conclude that $\phi^-_c$ and $\phi^+_c$ must satisfy,
\begin{equation}
	\label{eq:localization}
W[\phi^-_c] < 0
\ \ \ \mbox{and } \ \ \
W[\phi^+_c] > 0 .
\end{equation}
Finally, $\phi^-_c$ and $\phi^+_c$ must be dynamically reached at
$y=-\infty$ and $y=+\infty$, which according to (\ref{eq:phi}) is possible
{\it iff}:
\begin{equation}
	\label{eq:dynamics}
{\frac{d^2 W}{d\phi^2}}\Big|_{\phi^-_c} > 0
\ \ \ \mbox{and } \ \ \
{\frac{d^2 W}{d\phi^2}}\Big|_{\phi^+_c} < 0 ,
\end{equation}
or at least a similar condition for the first non-vanishing higher order
derivatives at $\phi^-_c$ and $\phi^+_c$.

\begin{figure}[htb]
\centerline{\epsfxsize=15cm\epsfbox{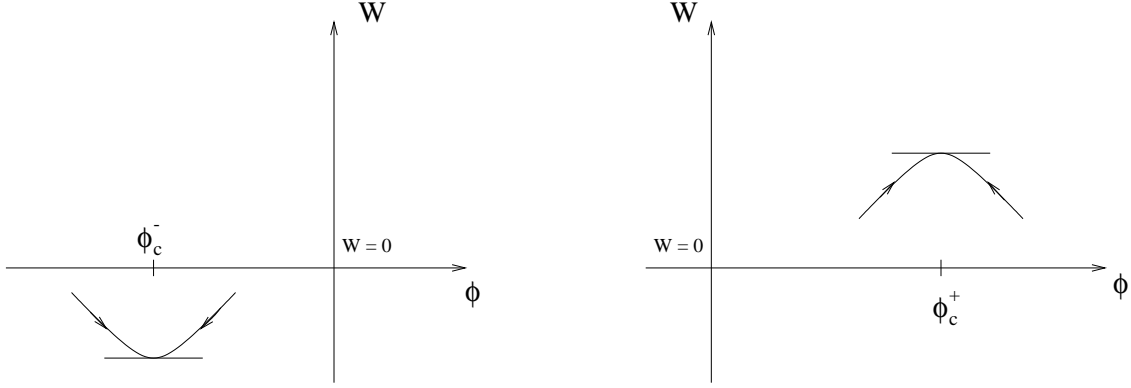}}
\caption{ Asymptotic behaviors of $W[\phi]$ leading to a singularity free
bulk geometry localizing gravity on the brane. The absence of singularities
is equivalent to the conditions: ${W^{'\,}}_c^- = {W^{'\,}}_c^+=0$;
the dynamics of the equations of motion require
${W^{''\,}}_c^- >0$ and ${W^{''\,}}_c^+<0$, while
the localization of gravity requires $W_c^- <0$ and $W_c^+ >0$.
}
\label{fig:W}
\end{figure}

Pictorially, the previous conditions are summarized in figure \ref{fig:W}.

Finally, the last equation of motion that relates the superpotential, $W$,
to the scalar potential in the bulk, $V$, partially fixes the possible values of
$\phi^-_c$ and $\phi^+_c$. Indeed this differential equation evaluated at infinity
gives:
\begin{equation}
	\label{eq:Vless0}
V[\phi^-_c] = -W[\phi^-_c]^2 < 0
\ \ \ \mbox{and } \ \ \
V[\phi^+_c] = -W[\phi^+_c]^2 < 0 .
\end{equation}
while a differentiation with respect to $\phi$ gives:
\begin{equation}
	\label{eq:V'=0}
{\frac{dV}{d\phi}}=
2 {\frac{dW}{d\phi}} \left(  {\frac{d^2W}{d\phi^2}} -W \right)
\ ; \ \mbox{ \ thus  } \ \ \
{\frac{dV}{d\phi}}_{|\phi^-_c} = 0
\ \ \mbox{and } \ \
{\frac{dV}{d\phi}}_{|\phi^+_c} = 0 .
\end{equation}

More information on $W$ can be obtained by considering higher order derivatives
of the differential equation between $W$ and $V$. Indeed it is easy to prove
by induction the following relation:
\begin{equation}
	\label{eq:Vn}
V^{(n)}
=
\sum_{k=1}^{n}
2 \binom{n-1}{k-1} W^{(k)} 
\left( W^{(n-k+2)} - W^{(n-k)} \right) \,
\end{equation}
where $V^{(n)}$ denotes the $n^{th}$ order derivative of $V$ and similarly for
$W^{(n)}$; in addition, we will denote by ${W^{(n)}}_c^\pm$ the values 
of $W^{(n)}$ at
$\phi=\phi_c^\pm$. At the second order, by evaluating (\ref{eq:Vn}) at
$\phi=\phi_c^\pm$, we obtain a quadratic equation for ${W^{(2)}}_c^\pm$:
\begin{equation}
2 {{W^{(2)}}_c^\pm}^2 - 2 W_c^{\pm} {W^{(2)}}_c^\pm - {V^{(2)}}_c^\pm
=0  \ .
\end{equation}
The superpotential will be real-valued provided that:
\begin{equation}
{V^{(2)}}_c^\pm > V_c^\pm/2 \ ,
\end{equation}
and then there are four different branches at each asymptotic point:
\begin{eqnarray}
&
W_c^\pm = \epsilon_1 \sqrt{-V_c^\pm} \ ,\\
&
{W^{(2)}}_c^\pm =  \epsilon_1 \frac{\sqrt{-V_c^\pm}}{2}
+ \epsilon_2 \frac{\sqrt{2 {V^{(2)}}_c^\pm - V_c^\pm}}{2} \ ,
\end{eqnarray}
with $\epsilon_1=\pm \epsilon_2=\pm 1$.

However the compatibility between the gravity localization requirement
(\ref{eq:localization}) and the dynamics of the differential equation
(\ref{eq:dynamics}) can be fulfilled {\it iff}
\begin{equation}
	\label{eq:Vpp}
{V^{(2)}}_c^\pm \geq 0 \ ,
\end{equation}
and only one out of the four branches is retained:
\begin{eqnarray}
\phi_c^- :
& W_c^- = - \sqrt{-V_c^-} \ \ \ \
{W^{(2)}}_c^- =  -\frac{\sqrt{-V_c^-}}{2}
+  \frac{\sqrt{2 {V^{(2)}}_c^- - V_c^-}}{2} \ ,
\\
\phi_c^+ :
& W_c^+ = \sqrt{-V_c^+} \ \ \ \
{W^{(2)}}_c^+ =  \frac{\sqrt{-V_c^+}}{2}
-  \frac{\sqrt{2 {V^{(2)}}_c^+ - V_c^+}}{2} \ .
\end{eqnarray}

At higher order, the relation (\ref{eq:Vn}) becomes linear in ${W^{(n)}}_c^\pm$
and allows to compute ${W^{(n)}}_c^\pm$ recursively in terms of lower derivatives:
\begin{equation}
	\label{eq:Wn}
{W^{(n)}}_c^\pm =
\frac{\scriptstyle
{V^{(n)}}_c^\pm -
{\displaystyle \sum_{k=3}^{n-1}}
2 \binom{n-1}{k-1} {W^{(k)}}_c^\pm
\left( {W^{(n-k+2)}}_c^\pm - {W^{(n-k)}}_c^\pm \right)
+2(n-1){W^{(2)}}_c^\pm {W^{(n-2)}}_c^\pm }
{\scriptstyle 2 \left( n{W^{(2)}}_c^\pm - W_c^\pm \right)} \ .
\end{equation}
Of course, for this expression to make sense for any integer $n>2$, it is important
that no solution to the equation $n{W^{(2)}}_c^\pm - W_c^\pm=0$ can be found.
However, the roots of this equation would satisfy:
\begin{equation}
\frac{{V^{(2)}}_c^\pm}{V_c^\pm}= \frac{2}{n} \left( 1- \frac{1}{n} \right) \ ,
\end{equation}
which is incompatible with the physical requirements
(\ref{eq:Vless0}) and (\ref{eq:Vpp}) of localized gravity.

At this stage, it is worth noticing that the uniqueness of a superpotential $W$ reaching a given
asymptotic point follows from our requirements of a localized gravity
without singularity. Otherwise, a scalar field $V$ in the bulk could
be constructed
such that the equation $n{W^{(2)}}_c^\pm - W_c^\pm=0$ has some solution in
which case
${W^{(n)}}_c^\pm$ would not be determined, leading to a continuum of solutions.

Finally, the whole expression of $W$ reaching the asymptotic points $\phi_c^\pm$
can be uniquely reconstructed from its derivatives through a Taylor expansion:
\begin{eqnarray}
y<0 :
&& W_- [\phi] = \sum_{n=0}^{\infty}
\frac{1}{n!} {W^{(n)}}_c^- (\phi-\phi_c^-)^n  \ ,
\\
y>0 :
&& W_+ [\phi] = \sum_{n=0}^{\infty}
\frac{1}{n!} {W^{(n)}}_c^+ (\phi-\phi_c^+)^n\ \ .
\end{eqnarray}
From the expression of the superpotential, we can solve the equations
of motion on the two sides of the brane. And the jump conditions are
given by:
\begin{equation}
\frac{W_+[\phi_0]-W_-[\phi_0]}{f[\phi_0]}
\ = \
\frac{W'_+[\phi_0]-W'_-[\phi_0]}{f'[\phi_0]}
\ = \ T \ .
\end{equation}
The first equation will fix the value, $\phi_0$, of the scalar field
on the brane while the second one fine-tunes the value of the brane tension.
Generically, we will obtain only discrete values of $\phi_0$. Indeed, if there
exist a continuum interval of solutions for $\phi_0$, then the jump
equation becomes a differential equation that can be integrated on this
interval and we obtain that $f[\phi]$ has to be proportional
to $W_+[\phi]-W_-[\phi]$, but in that case there is only one possible
value for the brane tension\footnote{\label{foot:f=W}
Whereas the non-canonically normalized
brane tension, $T$, is fine-tuned, the physical brane tension,
$T_{br}\propto f[\phi_0]T$, is not fine-tuned since the value of $\phi_0$
can vary continuously. Whether this is a solution to the cosmological
constant problem or not, beside the fine-tuning requires on $f[\phi]$,
depends whether SM loops will modify $T$ or $T_{br}$. This issue
deserves further analysis in future work.} In all the other cases, given two
asymptotic points, $\phi_c^\pm$, we will obtain discrete solutions for
$\phi_0$ and $T$. Moreover, using different critical asymptotic points, we usually find different values of
the brane tension and the number of values of $T$ that allows an infinitely
large extra dimension with localized gravity is related to the number of critical
points of the bulk potential $V$ satisfying \eqref{eq:Vless0},
\eqref{eq:V'=0} and \eqref{eq:Vpp}:
\begin{equation}
n_T \leq (3 n_C -2) n_S,
\end{equation}
where $n_T$ stands for the number of values of $T$ such that 
gravity is localized without a singularity, while $n_C$ is the number
of critical points of $V$ as defined above. 
The multiplicity factor comes from the fact that the scalar field
can asymptote either the same critical point or two adjacent ones on the two
sides of the brane. The upper bound has been semi-quantitatively corrected
by taking into account the average number, $n_S$, of solutions to the jump
equation for $\phi_0$. It may also happen that some values
of $T$ are degenerate. We will explicitly describe an example in the next
subsection.

It is worth noticing that in the case of an oscillatory bulk potential,
we will obtain an infinite number of discrete values for the brane tension
that is as quantized.

To complete the proof of the no-go theorem presented above, we need to 
show that there is no loophole in the above argument due to the fact that
we have used the superpotential formalism. The subtlety that one might 
worry about is that in the proof above we have implicitly assumed that
$\phi$ is monotonic, by writing the second order equations
(\ref{ein1})-(\ref{ein2}) in terms of first order equations involving $W$.
In particular, if $\phi$ is not monotonic (that is if $\phi'=0$ at a 
finite value of $y$) one does not have a globally defined superpotential
function $W(\phi )$, but instead one must define separate superpotential
functions $W_i$ for the regions between $y_i$ and $y_{i+1}$, where
$\phi' (y_i)=\phi'(y_{i+1})=0$. For these superpotentials that are not
globally defined it is then possible to have $W'(\phi_*)=0$ without satisfying
$V'(\phi_*)=0$.  However, it is impossible to continue the solution
beyond $\phi_*$. This by itself however may not be a problem, as long
as at $\phi_*$ one is smoothly switching over to another branch of 
$W$. Of course this switch-over can only happen at a point where $\phi'=0$, 
since otherwise $\phi$ is monotonic, and one can solve the equations in terms 
of the superpotential, which is well-defined around $\phi_*$. Next we show 
that such
possibilities do not get around the no-go theorem presented above. The
reason is that in order to have localized gravity with an infinitely
large extra dimension, we need $\phi'\to 
0, \phi'' \to 0$ for $|y|\to \infty$, and therefore we find from
the second order equation (\ref{ein3}) that $V'\to 0$. Thus the
critical points at infinity must belong to an ``ordinary branch'' described
above, where $W$ can be continued at both sides of $\phi_c$. However,
once we are on an ``ordinary branch'' which can be globally defined,
all the critical points will actually happen at $V'=0$. Since the
only possibility for switching over to another branch is at $W'=0$, and
at those points $V'=0$, one can never switch off the ordinary branch, and
therefore one can not circumvent the no-go theorem by gluing non-monotonic
$\phi$'s together.

%%%%%%%%%%%
\subsection{A numerical example}

\begin{figure}[htb]
\centerline{\epsfxsize=12cm\epsfbox{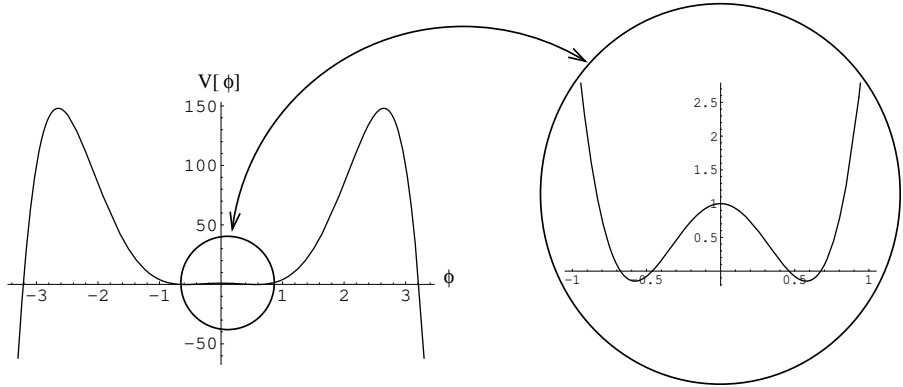}}
\caption{ The scalar bulk potential \eqref{V1}. The physically interesting region, near
the two negative stationary points, is emphasized.}
\label{fig:V2}
\end{figure}

In this section we demonstrate the ideas of the previous sections in a 
numerical example.  The bulk potential in this example (Fig.~\ref{fig:V2}) is
\begin{equation}
	\label{V1}
V(\phi)=-\phi^6+11\phi^4-7\phi^2+1,
\end{equation}
which is generated by the superpotential $W[\phi]=\phi^3-\phi$.

As in some of our previous illustrative examples the potential
\eqref{V1} is unbounded from
below, and so the theory might be unstable to quantum fluctuations.  
However, we will only
be concerned with static solutions and for our purposes any instabilities
will not be relevant.

As emphasized in Figure \ref{fig:V2}, the bulk potential \eqref{V1} has two negative
stationary points at $\phi=\pm 1/\sqrt{3}$. According to our no-go theorem,
there are isolated superpotentials solving the equations of motion with 
critical
points at $\pm 1/\sqrt{3}$. Here these superpotentials are very simple
because from \eqref{eq:Wn} only a finite number of derivatives are
non-vanishing thus the superpotentials are polynomial and take
the form
\begin{eqnarray}
W[\phi] = \phi^3-\phi & \mbox{reaches } \phi_c=\pm 1/\sqrt{3} \ \
\mbox{on the negative (positive) branch},
\\
W[\phi] = -\phi^3+\phi & \mbox{reaches } \phi_c=\pm 1/\sqrt{3} \ \
\mbox{on the positive (negative) branch}.
\end{eqnarray}
Depending on which critical point we want to asymptote at infinity, we can construct
four types of solutions that localizes gravity
\begin{itemize}
\item for $(\phi_c^+= 1/\sqrt{3}, \phi_c^-= 1/\sqrt{3})$~: the solution is
\begin{eqnarray}
&& \phi (y) = \sfrac{1}{\sqrt{3}}\tanh (\sqrt{3}|y-y_c^\pm|) \\
&& A (y) = \sfrac{1}{18} \tanh^2 (\sqrt{3}(y-y_c^\pm)) +
\sfrac{2}{9} \ln \cosh (\sqrt{3}(y-y_c^\pm)) + a_\pm
\end{eqnarray}
where $y_c^\pm$ and $a_\pm$ are four constants of integration to be determined
by the continuity and jump conditions. The continuity conditions
imply that $y_c^+=-y_c^-$ and $a_+=a_-$. The jump equations will depend
on the precise form of $f[\phi]$.
Except in the degenerate case where $f[\phi] \propto W_+[\phi]-W_-[\phi]$
that has been discussed in a footnote on page \pageref{foot:f=W}, we will
generically obtain a finite number of discrete values for $\phi_0$.
For instance in the case of an exponential coupling,
$f[\phi]=a\, e^{b\phi}$, the values of $\phi_0$ will satisfy
\begin{equation}
\frac{3\phi_0^2-1}{\phi_0(\phi_0^2-1)} \  = \ b \ ,
\end{equation}
which admits three solutions whatever the value of $b$ is. And thus
there exist three values for the brane tension that will lead
to a localized gravity.
\item  for $(\phi_c^+= 1/\sqrt{3}, \phi_c^-= -1/\sqrt{3})$
or $(\phi_c^+= -1/\sqrt{3}, \phi_c^-= 1/\sqrt{3})$~:
in those cases, there is no discontinuity in the superpotential and the
brane tension has to vanish.
\item for $(\phi_c^+= -1/\sqrt{3}, \phi_c^-= -1/\sqrt{3})$~:
this case is analogous to the first one and, for an exponential coupling,
three values for the brane tension are possible and they are just the opposite
of the ones obtained in the first case.
\end{itemize}
%
%%%%%%%%%%%%%%%%%%%%%%%

%
\begin{figure}[htb]
\centerline{\epsfxsize=15cm\epsfbox{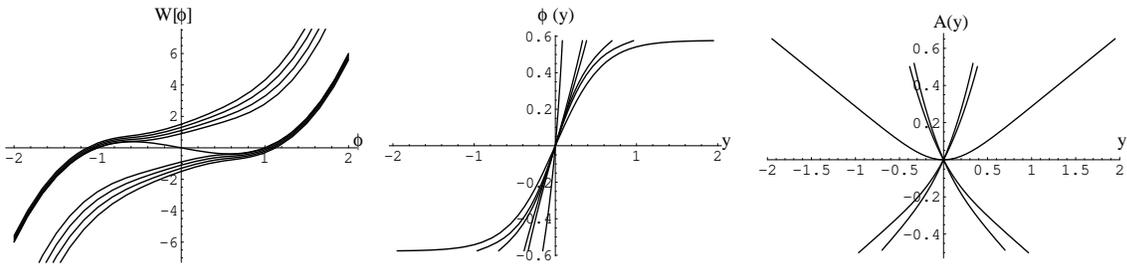}}
\caption{Solutions for $W[\phi]$, $\phi(y)$ and $A(y)$ in the bulk potential
\eqref{V1}.  There is a unique, fine-tuned, regular solution which approximates
the Randall-Sundrum solution for large $|y|$.  All other solutions are singular.}
\label{fig:dzf}
\end{figure}

Besides the previous fine-tuned solutions with an infinitely large extra dimension
and localized gravity, all other solutions will either decouple gravity on
the brane or involve a horizon at a finite distance. We would like now to
examine numerically the self-tuning of these singular solutions. To do that
we scan for solutions to \eqref{V}
by fixing the value of $W[\phi]$ at $\phi=0$. The solutions are plotted in
Fig.~\ref{fig:dzf}.  Note that in agreement with the no-go theorem of the
previous section there is a unique solution in this branch of solutions
with stationary points at the positions of the minima of $V[\phi]$.  As a
result of the uniqueness of the solitonic solution, there are no solutions
(in this branch) with $0<|W[0]|<W[\phi_\ast]\simeq .385$, where $\phi_\ast$ is 
the value of the field at the position of the local minimum of $V[\phi]$, in
this case $\phi_\ast\simeq -.577$.  Also in agreement with the results of
previous sections, the solitonic solution is the unique regular solution for 
$\phi(y)$ and $A(y)$, which in this case approximates the Randall-Sundrum
solution for large $|y|$.

\begin{figure}[htb]
\centerline{\epsfxsize=8cm\epsfbox{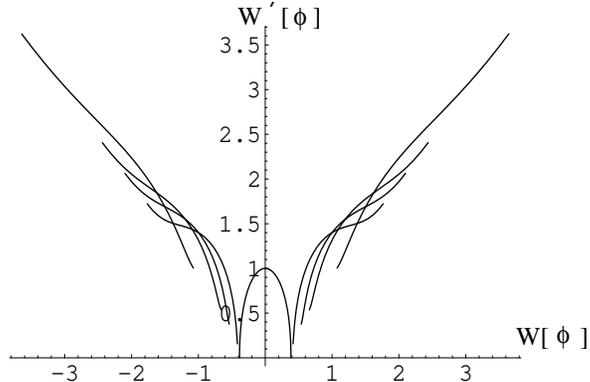}}
\caption{Parametric plot of several solutions for $(W[\phi],W'[\phi])$, which
specify the boundary conditions at the wall.}
\label{fig:medusa}
\end{figure}

In order to determine the range of boundary conditions at the brane which
could be satisfied, we note that except for the solitonic solution, all
other solutions span an infinite range in $\phi$.  As noted in 
Section~4.2 the boundary conditions can be expressed in terms of
$W[\phi]$ and $W'[\phi]$ at the boundary.  Hence, if there is a space-filling
range of solutions $(W[\phi],W'[\phi])$
then no fine-tuning is necessary in order to have a static solution with
arbitrary boundary conditions in that range.  In other words, a given
$(2W[\phi_0],2W'[\phi_0])$ can be equated with a certain (orbifold) boundary 
condition
$(f[\phi_0]\,T,f'[\phi_0]\,T)$. Then, given one such $(f[\phi_0]\,T,
f'[\phi_0]\,T)$
there is a continuous set of solutions around that point.  That means that
for a given $(f[\phi_0],f'[\phi_0])$ in a certain range of $f'/f$ there is
a range of tensions $T$ for which there is a solution; hence the theory is
self-tuning.
Fig.~\ref{fig:medusa} is
a parametric plot of $W[\phi]$ versus $W'[\phi]$ for several solutions of
$W[\phi]$.  It is intended to illustrate the fact that there is a space-filling
region for $W > .385$.  Given any boundary condition parametrized by
$f[\phi]$ and $T$, $T$ can be rescaled by an amount given by the
intersection of the set of solutions $(2W,2W')$ with a line through the origin
and $(fT,f'T)$.  As is generically expected, one can see from 
Figure~\ref{fig:medusa} that if $f,f'$ are ${\cal O}(1)=
{\cal O}(M_5)$, then $T$ can be rescaled by ${\cal O}(M_5)$ without
eliminating a solution.  Hence we have demonstrated
the self-tuning of this model.  However, the caveat is that because of the
isolated solitonic solution, there is a fine-tuned region near $W,W'\sim
{\cal O}(1)$.  If $f$ is ${\cal O}(1)$ at some matching point $\phi_0$,
and $T$ is ${\cal O}(M_{EW}/M_5)\ll 1$ then a fine tuning is reintroduced
because of the uniqueness of the solitonic solution.  There is still
a large region of parameter space where the theory is self-tuning, but
whether that region is natural or not requires exploration.    This phenomenon
is a result of the existence of solitonic solutions, and is nongeneric.
Furthermore,
if we do not require orbifold boundary conditions, then once again self-tuning
is natural.
Note also that asymptotically the parametric
plots of $W$ vs. $W'$ include the line $W=W'$.  This is generic in any region
where the solutions satisfy $W[\phi]\gg V[\phi]$.  This also implies
that for any solution $f\sim f'$, there is a very large range for $T$ for
which there are solutions.  This behavior for large values of $W$ is
common, and extends the range of $T$ over which self-tuning occurs possibly
to $\pm\infty$ if $f[\phi]=f'[\phi]$, which may be natural from a
stringy perspective \cite{ADKS,KSS}.

%%%%%%%%%%%%%%%%%%%%%%%%%%%%%%%%%%%%%%%%%%%%%%%%%%%%%%%%%%%%%%%%%%%%%%%%%%%%%%
\section{Resolution of singularities and fine-tuning}
\setcounter{equation}{0}
\setcounter{footnote}{0}

In this section we reconsider the resolution of singularities\footnote{
As they stand in our solutions, Einstein's equations are not
satisfed at the singularity but require a singular stress-energy tensor
located at the singularity that may correspond to a brane: the introduction
of this brane, for instance, is what we mean by resolution of the
singularity.} as proposed in \cite{deAlwis,nilles}.
As we have seen, the case of zero bulk potential, which
is the case studied in \cite{nilles}, is quite non-generic.  There is a
shift symmetry in the scalar field which makes the boundary conditions
more constraining, and two of the boundary conditions turn out to have the
same form.  Hence, we study the generic case, and propose a new resolution
of the singularity which may restore self-tuning.

The idea of \cite{deAlwis,nilles} is to add a brane at each of the
singularities such that the
equations of motion, or boundary conditions, are satisfied there, as well.  
As pointed out in \cite{deAlwis,nilles}\footnote{See
also the first reference in \cite{brane-cc-new}
for a discussion on vanishing 4D effective vacuum energy.}
the singularities contribute to the effective 4D energy density
when integrated over the extra dimension, and the contribution of the
singularities is essential for vanishing of the 4D cosmological constant.

However, addition of branes at the singularities adds new boundary conditions.
In order to answer the question of whether or not the theory is self-tuning
we must better understand the continuation past the singularities.  In
\cite{nilles} it is proposed that the spacetime is either periodically
continued or cut-off at the singularity.  In the case that the spacetime
is cut-off at the singularities on each side, there are generically two new 
boundary 
conditions at each singularity, but no additional free parameters.  Hence,
without orbifold boundary conditions there are $3+3-1=5$ free parameters
and $4+2+2=8$ boundary conditions, and the system is overconstrained.
If one
imposes orbifold boundary conditions, then there are only half as many 
additional boundary conditions 
($2+2=4$ boundary conditions in all), but only $3-1=2$ free parameters. In
either case the system is overconstrained.  Hence, a fine-tuning is required.
Although in the absence of a bulk potential the situation is modified because
of the non-generic features mentioned above, in that case a fine-tuning
at each of the singularities is required, as well.

\begin{figure}[htb]
\centerline{\epsfxsize=7cm\epsfbox{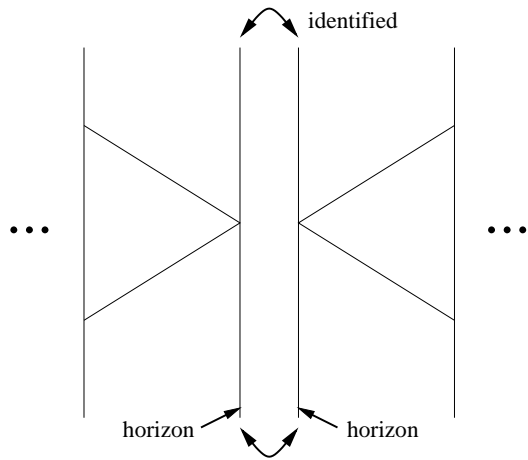}}
\caption{Penrose diagram for causally disconnected regions glued together
at the resolved singularity.}
\label{fig:penrose}
\end{figure}

Although the details of any continuation of the spacetime beyond the 
singularity will depend on
quantum gravitational dynamics, we propose another scenario which
does slightly better than the cut-off scenario, although
fine-tuning will still be required.  The
singularities are generically horizons ({\em c.f.} Sections~\ref{sec:exp}) because the warp factor $e^{-2A(y)/(D-1)}$ vanishes there.
Hence, light does not cross the horizon and we can imagine a scenario in which
there are bulk fields living beyond the singularity, out of causal contact
with our universe.  The Penrose diagram for this scenario is illustrated in 
Figure~\ref{fig:penrose}.  However, the bulk fields across the horizon
would provide an additional three free parameters which one might hope 
would help avoid fine-tuning.  More precisely, following the counting in
Section~\ref{sec:self-tune}, if there are orbifold boundary conditions,
then there are $3+3-1=5$ free parameters from the bulk fields on both sides
of the singularity, but $2+4=6$ boundary conditions.  Hence, although the
system is less constrained than the case in which the spacetime is cut-off
at the singularity, one fine-tuning is still required.

One might suspect
that additional fields would add additional degrees of freedom which could
help in self-tuning. 
For example, the addition of a scalar field with second derivatives in its
equation of motion would contribute two additional free parameters on each
side of a brane, but only two boundary conditions (continuity and change
in the derivative at the brane).  Hence in a system of branes there are net, at
least naively, two additional free parameters from the extra scalar field,
which could be used to restore self-tuning at the singularities or perhaps
even produce nonsingular solutions.  However, a more detailed analysis is 
required in this case.

%%%%%%%%%%%%%%%%%%%%%%%%%%%%%%%%%%%%%%%%%%%%%%%%%%%%%%%%%%%%%%%%%%%%%%%%%%%%%%
\section{Conclusions}

We have studied brane worlds coupled to a scalar field and have found that
self-tuning is a generic feature of these models.  In these models the
dynamics of the scalar field provides additional degrees of freedom, which
generically alleviates the need for fine-tuning of static
solutions.  We have reexamined the exactly solvable models, two of which
were studied previously \cite{ADKS,KSS}, and have found that those case
are more constrained than the generic case because of a shift symmetry
in the scalar field in these models.  Still, these theories are self-tuning,
including the case of an exponential potential.  Whereas in \cite{KSS} a
fine-tuning was necessary in this case due to a particular ansatz for the
scalar and graviton fields, we showed that the more general solution is not
fine-tuned, in agreement with a counting of free parameters in these models.
We demonstrated that singularities in the self-tuned solutions are 
generic if gravity is to be localized, and we presented a no-go theorem to this
effect.  We provided perturbative and numerical techniques in order to
calculate the self-tuned solutions, and we illustrated the major points 
of the paper via several numerical examples.  Finally, we pointed out that 
the fine-tuning that is required in order to resolve the singularities in the
spirit of \cite{deAlwis,nilles} for the case of zero bulk potential is generic.

%%%%%%%%%%%%%%%%%%%%%%%%%%%%%%%%%%%%%%%%%%%%%%%%%%%%%%%%%%%%%%%%%%%%
\section*{Acknowledgments}
We are extremely grateful to Chris Kolda for collaboration at early
stages of this project. We thank Nima Arkani-Hamed, Tanmoy
Bhattacharya, Pierre Bin\'etruy, Martin Schmaltz and Anupam Singh for useful
discussions. C.C. thanks the Theory Group at Berkeley for hospitality
while this work was initiated. C.C. is a J. Robert Oppenheimer fellow
at the Los Alamos National Laboratory. C.C., J.E. and T.H. are 
supported by the US Department of energy under contract W-7405-ENG-36.
C.G. is supported in part by the US Department of energy under Contract
DE-AC03-76SF00098 and in part by the National Science Foundation under
grant PHY-95-14797.

%%%%%%%%%%%%%%%%%%%%%%%%%%%%%%%%%%%%%%%%%%%%%%%%%%%%%%%%%%%%%%%%%%%%%%

\end{document}